\documentclass[pre,twocolumn,showpacs,amsmath,amssymb,floatfix]{revtex4}

\usepackage{graphicx}

\usepackage{times}

\begin{document}
\def\OMIT#1 {{}}   
\def\secc#1{{\it #1} ---}
\def\subsecc#1{{\it #1} ---}
\def\MEMO#1 {{}}   

\newcommand{\dagg}{^\dagger}
\def \Tr {{\rm  Tr}}
\def \la {{\langle}}
\def \ra {{\rangle}}
\def \eqref#1{(\ref{#1})}
\def \rr {{\bf r}}
\newcommand{\gop}{\hat{g}}
\newcommand{\hop}{\hat{h}}
\newcommand{\Xop}{\hat{X}}
\newcommand{\Yop}{\hat{Y}}
\newcommand{\Pop}{\hat{P}}
\newcommand{\Qop}{\hat{Q}}
\newcommand{\rhoop}{\hat{\rho}}
\newcommand{\rhoC}{\hat{\rho}^C}


\title{Correlation density matrix:  an unbiased analysis of exact diagonalizations} 

\author{Siew-Ann Cheong} 
\altaffiliation[Current and permanent address: 
Div. of Physics and Appl. Physics, School of Physical and Mathematical
Sciences, Nanyang Technological Univ.,
1 Nanyang Walk, Block 5, Level 3, Singapore 637616, Rep. of Singapore]
{}
\author{Christopher L. Henley}
\affiliation{Laboratory of Atomic and Solid State Physics, Cornell University,
Ithaca, New York, 14853-2501}

\begin{abstract}
Given the ground state wavefunction for an interacting lattice model,
we define a ``correlation density matrix'' (CDM) for 
two disjoint, separated clusters $A$ and $B$, to be the density
matrix of their union, minus the direct product of their
respective density matrices.
The CDM can be decomposed systematically
by a numerical singular value decomposition, to provide
a systematic and unbiased way to identify the operator(s)
dominating the correlations, even unexpected ones.
\end{abstract}

\pacs{02.70.-c, 71.10.Pm, 71.10.Hf}


\maketitle

\OMIT{Grand summary:
The density matrix (DM) of the union $A\cup B$ of separated
clusters A and B, minus the direct project of the DM's of A and
B, is defined as the ``correlation DM''.  Its operator norm
measures all correlations (even unexpected ones) between A and B.
Using singular-value decomposition we write
   \begin{equation}
   \rhoC =
    \sum _\nu \sigma_\nu \Xop_\nu(A) \Yop_\nu(B) ,
   \end{equation}
where $\Xop_\nu$ and $\Yop_\nu$ are normalized operators on
the respective clusters; the terms represent different correlation
functions, which are naturally ordered by the magnitudes $|\sigma_\nu|$.
This provides a systematic, unbiased numerical method to
identify the important correlations, given the ground state
wavefunction.  The procedure is tested (see below) on ladders of
spinless fermions with infinite nearest-neighbor repulsion.}


The ground state of a strongly-interacting, quantum-mechanical lattice model 
(with spin, boson, or fermion degrees of freedom) is characterized by 
long-range order, power-law correlations, or the lack of these.
When such a system is studied numerically, it may be unclear
{\it a priori} what kind of correlation will be dominant
-- especially in cases where exotic order or disorder
are possible, such as the doped square-lattice Hubbard model, or
(better) the highly frustrated $s=1/2$ Kagome antiferromagnet;
in the latter system spin-spin, spin-Peierls, spin-nematic, or 
chiral order parameters were all serious candidates~\cite{kagome}.
Before computing the ground state correlations, 
one must first guess which operators are important --  
a choice which is necessarily biased by one's prior knowledge or 
preconceptions, and is problematic for hidden or exotic orders.


In contrast, approaches based on the density matrix (DM)
of a cluster of several sites are unbiased -- apart from 
specification of that cluster --  since
the DM specifies the expectation of every operator 
local to the cluster -- including the ``key operator(s)''
meaning those having long range order (i.e. order parameter)
or having strong correlations.
For exact diagonalizations (ED) of interacting systems, the DM was used 
as a diagnostic to compare different system sizes~\cite{SAC-interacting}
or truncations of the Hilbert space \cite{Cap04a}.


Here we propose a new application of the density matrix as 
a way to uncover correlations/orders from numerics
{\it without} requiring any foreknowledge of what kinds to expect.
Consider two small disjoint clusters $A$ and $B$
(identical apart from a translation),
either cluster having a Fock-Hilbert space of dimension $D$.
\OMIT{call the clusters' union $A\cup B$ the ``supercluster''}
Let $\rhoop^{AB}$ be the many-body density matrix for 
the disconnected ``supercluster'' $A\cup B$, constructed from
the whole system's ground state wavefunction by tracing out all other sites,  
with $\rhoop^A$ and $\rhoop^B$ similarly defined.
Then we define the {\it correlation density matrix} (CDM) to be
    \begin{equation}
    \label{eqn:correlationdm}
           \rhoC \equiv \rhoop^{AB} - \rhoop^A\otimes\rhoop^B.
    \end{equation}
\OMIT{where $\rhoop^A\otimes\rhoop^B$ is the direct product.}
If there were no correlations between clusters $A$ and $B$,
then  $\rhoop^{AB}=\rhoop^A\otimes \rhoop^B$ and $\rhoC=0$.  

The CDM defined in \eqref{eqn:correlationdm} contains
all possible inter-cluster correlations~\cite{FN-entangle}.
Write the (``connected'') correlation of the fluctuations of 
any two operators as $\la \Pop \Qop \ra_c \equiv
\la \Pop \Qop \ra - \la \Pop \ra \la \Qop \ra$; 
then if $\Pop(A)$ and $\Qop(B)$ act on clusters $A$ and $B$,
    \begin{equation}
    \label{eqn:CPQ-CDM}
         \la \Pop(A) \Qop(B)\ra _c =
           \Tr \big[ \rhoC \Pop(A) \Qop(B) \big] .
    \end{equation}

\secc{Index relabeling and the operator singular-value decomposition}
The key notion underlying our processing of the CDM is,
given the $D\times D$ matrix representing an operator 
on a cluster's $D$ dimensional Hilbert space, to rewrite
it as an $D^2$-component {\it vector} of complex numbers
using fused indices~\cite{FN-partialtransp}
$(a',a) \leftrightarrow \alpha(a',a), \, (b',b)\leftrightarrow \beta(b',b)$.
Say that $\rhoC$ is known in terms of the product states
$|a'\ra |b'\ra$ and $|a\ra  |b\ra$ of the occupation-number
basis on the clusters~\cite{FN-fermionsign}.
Then
   \begin{equation}
      \label{eqn:corrdmfused}
      \rhoC = \sum_{a',b',a,b} \rhoC_{a'b',ab} |a'\ra |b'\ra \la a| \la b|
            \equiv
        \sum _{\alpha\beta} C_{\alpha\beta} \; \gop_\alpha\, \hop_\beta
   \end{equation}
where $\rhoC_{a'b',ab} \equiv C_{\alpha(a',a),\beta(be,b)}$.
Here $\gop_\alpha\equiv |a'\ra \la a|$ and $\hop_\beta \equiv |b'\ra \la b|$ 
are bases for the respective clusters $A$ and  $B$, 
manifestly orthonormal in terms of the \emph{Frobenius norm} 
   \begin{equation}
       \|\Pop\|_F^2 \equiv \sum_{a',a} |P_{a',a}|^2 =
       \Tr \left( \Pop\dagg \Pop\right)
   \label{eqn:frob-norm}
    \end{equation}
for any operator $\Pop$, and the {\it Frobenius inner product}
   \begin{equation}
        \label{eqn:frob-innerprod}
        (\Pop,\Qop)_F \equiv \sum_{a',a} P_{a',a}^* Q_{a',a} =
       \Tr \left(\Pop\dagg \Qop\right) .
   \end{equation}
(In the fused-index notation, Eqs.~\eqref{eqn:frob-norm}) 
and \eqref{eqn:frob-innerprod} take on the usual form of a vector norm
and vector inner product.)


Next a numerical singular value decomposition  can be
made of $C_{\alpha\beta}$ as a matrix of complex numbers:
    \begin{equation}
    \label{eqn:corrdmmatrixsvd}
           C_{\alpha\beta} = \sum _\nu 
                  \sigma _\nu U_{\nu\alpha} V_{\nu\beta}
    \end{equation}
where $U$ and $V$ are unitary matrices,  and
$\{ \sigma_\nu: \nu =1, \ldots, D^2 \}$ are the singular values.
[Eq.~\eqref{eqn:corrdmmatrixsvd}
can also be written in the matrix form $C = U^T\Sigma V$,  where
$\Sigma \equiv {\rm diag}( \{ \sigma_\nu \} )$.]
Substituting  \eqref{eqn:corrdmmatrixsvd} into \eqref{eqn:corrdmfused}, we
obtain the {\it operator singular-value decomposition},
      \begin{equation}
      \label{eqn:operatorsvd}
           \rhoC = \sum _{\nu=1}^{D^2}
                         \sigma_\nu {\Xop_\nu} (A) \Yop_\nu (B)
      \end{equation}
\OMIT{where $D^2$ is the dimension of the operator algebra on either cluster.}
This (simple but powerful) expression is the key formula of our paper.
\OMIT{(All results apart from \eqref{eqn:operatorsvd} were
independently discovered by A. L\"auchli.)}
Each term represents the correlated quantum fluctuations 
of Frobenius-orthonormalized basis operators~\cite{FN-traceXY},
$\Xop_\nu \equiv \sum _\alpha U_{\nu\alpha} \gop_\alpha$
on cluster $A$ 
and $\Yop_\nu \equiv \sum _\beta V_{\nu\beta} \hop_\beta$,
on cluster $B$.
\OMIT{{\it independent} of the operators with other indices.}

Recalling \eqref{eqn:CPQ-CDM}, we can rewrite any correlation
    \begin{equation}
      \la \Pop(A) \Qop(B)\ra_c = 
          \sum _\nu \sigma_\nu (\Xop_\nu\dagg,\Pop(A))_F 
           (\Yop_\nu\dagg, \Qop(B))_F
    \end{equation}
in terms of Frobenius inner products \eqref{eqn:frob-innerprod}.
In particular, 
$\la \Xop_\nu(A)\dagg \Yop_\tau(B)\dagg\ra_c = \sigma_\nu \delta_{\nu\tau}$,
Thus $\{ \Xop_\nu(A)\dagg \}$ and $\{ \Yop_\nu (B)\dagg \}$ are
the natural bases into which operators $\Pop(A)$ and $\Qop(B)$
should be decomposed.
Each $|\sigma_\nu|$
is a normalized measure of the strength of the corresponding 
inter-cluster ground state correlation.
\OMIT{of $\Kop_\nu$  and $\Lop_\nu$.}
By convention, we order the
singular values $\sigma_1 \geq \sigma_2 \geq \cdots \geq \sigma_{D^2} \geq 0$.
This ordering gives a means of approximating 
$\rhoC$ by retaining just the first few terms in the expansion
\eqref{eqn:operatorsvd}.

\OMIT{
The rows of matrix $U_{\alpha\nu}$ are left eigenvectors.
The matrix elements of operator $\Xop_\nu$
are $U _{\alpha\nu} \to U_{a'a, \nu}$ in the original indices.}

\begin{figure}
\includegraphics[width=7.0cm,angle=0]{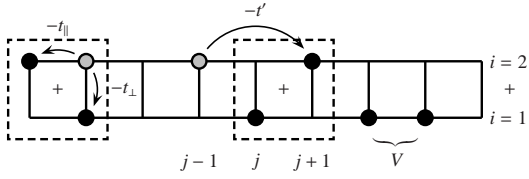}
\caption
{Model: spinless fermions, with hardcore excluding nearest-neighbors, 
on a ladder, with longitudinal hopping $t_\parallel\equiv 1$, transverse
hopping $t_\perp$, and correlated hopping $t'$.
The correlation density matrix involves two clusters, each of 
$2\times 2$ sites, with their centers (marked +) separated by $r$.
This ladder has length $L=8$, with periodic boundary conditions
as indicated by the $+$ at right edge.
\OMIT{(SAC)
Should we add (b) small figure with the rough phase diagram?}
}
\label{fig:ladder}
\end{figure}

Observe that $\|\rhoC\|^2= \sum _\nu |\sigma_\nu|^2$ is a 
basis-invariant measure of the total correlations between $A$ and $B$.  
\OMIT{It is zero when $\rhoop^{AB} \equiv \rhoop^A \otimes \rhoop^B$.}
Since~\cite{FN-rhoCbound} 
    \begin{equation}
       \|\rhoC\|_F^2= \|\rhoop^{AB}\|_F^2 - 
       \|\rhoop^{A}\|_F^2  \; \|\rhoop^{B}\|_F^2,
    \label{eqn:rhoCbound}
    \end{equation}
it follows that $\|\rhoC\|_F^2 \leq 1 - 1/D^2 \approx 1$, 
which gives a standard of comparison for numerically obtained $\sigma_\nu$'s.


The CDM typically inherits various symmetries from the input
wavefunction (ultimately from the Hamiltonian), such as 
spin-rotations, lattice rotations/reflections, or fermion number 
conservation~\cite{FN-restoresymm}.
The matrix $C_{\alpha\beta}$ breaks up into symmetry-labeled blocks, 
which (as with diagonalization) can be singular-value-decomposed
independently.  Each term in the expansion \eqref{eqn:operatorsvd} 
is thus assigned to a sector according to the
quantum numbers carried by $\Xop_\nu$ and $\Yop_\nu$, and each
sector is interpreted as representing a different kind of orrelation.

A convenient test bed to study CDM properties is
a non-interacting system (including BCS states)
for which density matrices can be calculated exactly,
~\cite{free-fermion-DM}.
We analytically checked the CDM and its operator SVD for a 
free Fermi sea in one dimension (Ref.~\onlinecite{SAC-thesis}, 
chapters 5 and 6), finding the expected FL correlations 
with an $r^{-1/2}$ envelope and CDW correlations with an $r^{-2}$
envelope.


\secc{ Ladder model: limiting regimes and operator classes}
We now test the CDM method on a toy system (Fig.~\ref{fig:ladder})
in which spinless fermions hop on a two-leg ladder of length $L$;
they are forbidden to occupy adjacent sites (i.e., the
nearest-neighbor repulsion is $V=\infty$).  
Three kinds of hopping amplitudes  appear:  $t_\parallel\equiv 1$ along
legs, $t_\perp$ along rungs, and $t'$ a ``correlated
hop'' conditioned on a second fermion, 
  $ - t' (c\dagg_j c_i + c\dagg_i c_j) \hat{n}_k$;
here $i$, $j$ are two steps apart on the same leg, and
$\hat{n}_k$ is the number operator for the site
between $i$ and $j$ on the opposite leg 
(which would block the $t_\parallel$ hops).

\begin{figure}
\includegraphics[width=8.5cm,angle=0]{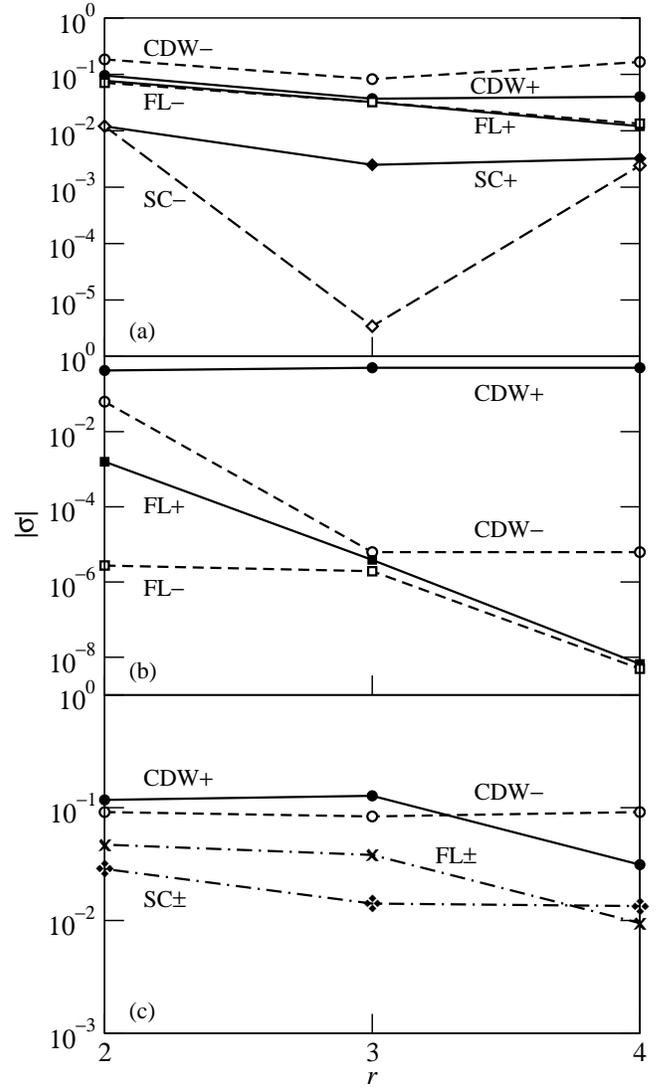}
\caption
{
Each plot shows (on a log scale) the magnitude of the
largest singular value for each symmetry sector of operators. 
The symmetries are labeled ``CDW'' for number operator
(or any combination $c_i\dagg c_j$ in the same cluster);
``FL'' for single creation/annihilation (i.e. the correlation
function  is a 2-point Green's function); ``SC'' for
superconducting (combination $c_i\dagg c_j\dagg$ in same cluster).
The symmetry label $\pm$ denotes even/oddness under
exchanging the legs of the ladder.  
In every case, there are 4 particles on a ladder of length
$L=8$, and twist boundary condition averaging was used.
(a). 
\OMIT{Thesis Fig. 8.74; or is 8.75 better?}
No-passing ladder with $t_\perp=0.1$, $t'=0$;
\OMIT{with fermions on alternating legs}
(b). 
 \OMIT{Fig. 8.77}
Rung-fermion case (each fermion delocalized on a rung)
with $t_\perp=100$, $t'=0$;
SC singular values do not appear since they are 
$\sim 10^{-15}$.
(c). Boson pair state: $t_\perp=0, t'=100$.
 \OMIT{Fig. 8.78}
}
\label{fig:sval-plots}
\end{figure}

The phase diagram 
(see Ref.~\onlinecite{SAC-thesis}, Fig. 8.1)
may be understood through the three limiting cases in which one hopping 
dominates. \OMIT{separated by transitions or crossovers}
(a) $t_\parallel$ dominant (``no-passing'' limit):
the leg index is a conserved flavor; the model
reduces to a free fermion chain (with fermions on alternate legs) 
(b) $t_\perp$ dominant (``rung-fermion'' limit):
each fermion delocalizes on a rung, so at low energy the 
model maps to reduces to a fermion chain with nearest neighbors excluded; 
(c) $t'$ dominant (``paired'' limit):
fermions bind into effective ($p$-wave) boson pairs
(in one dimension, with nearest neighbors excluded).
\OMIT{There are two flavors according to whether the fermion on
leg 1 is on even or odd sites.}
Regime (c) must be dominated by superconductivity at large length scales.

Each of the three limiting cases maps nontrivially to free fermions. 
Elsewhere~\cite{SAC-IPE} we
derived from these maps a semi-analytic method (``intervening
particle expansion'') to calculate various correlation functions;
the results of Ref.~\onlinecite{SAC-IPE}
have illuminated the present calculation.
The asymptotic behaviors (as expected) are that of a Luttinger liquid: 
power-law decays, with possibilities of commensurate locking 
when the filling is a rational fraction.

We performed exploratory exact diagonalizations
using periodic boundary conditions,
with four fermions on a ladder of length $L=8$,
the smallest (nontrivial) case at $1/4$ filling.
(This is the most interesting filling --
and the hardest, since the Hilbert space is largest
at filling 1/4: see Ref.~\onlinecite{zhang}(b), appendix.)
The largest block matrix for a sector is $27 \times 27$.
(As in our earlier ED studies on the square lattice~\cite{zhang,SAC-interacting},
the spinlessness and the neighbor exclusion greatly limit the Hilbert space
compared to e.g. a Hubbard system of the same dimensions.)
To minimize finite-size effects on the density matrices, it 
was necessary to use phase-twist boundary conditions~\cite{twist}
 (i.e. to
thread flux through the ``ring'' of sites) and average over
21 distinct phase angles.  (See Ref.~\onlinecite{SAC-interacting}
and Sec. 8.2.4 of Ref.~\onlinecite{SAC-thesis}).
\OMIT{At the system sizes accessed by ED,  finite-size effects (especially
shell-filling effects) are in general daunting~\cite{SAC-interacting}.  
Our remedy of twist-boundary condition averaging, 
demands tedious repeated diagonalizations with
different phases imposed on the boundary conditions.
Each parameter point (for $L=8$ with 11 twist angles) 
would take $\sim 4$ days on a 2.0-GHz workstation with 1 gigabyte memory.
Part of the slowness is that (i) the code was written in Octave
(ii) we preformed complete (not Lanczos) diagonalization
(iii) it couldn't be parallelized (this workstation has just
two processors).
(iv). Every wavevector sector was tried, since we did not
know {\it a priori} which one contains the ground state.)}


Each of our two clusters is $2\times 2$ (two adjacent rungs)
as shown in Fig.~\ref{fig:ladder}, the smallest cluster that  can capture
superconducting correlations; each cluster's Hilbert space has 
dimension $D=7$.
The operators $\{ \Xop_\nu , \Yop_\nu \}$,
emerging from the operator singular-value decomposition,
are classified into three main categories, according
to the fermion number change $\Delta F$ they carry:
(i) CDW (charge-density-wave-like), $\Delta F=0$,
e.g. the number operator $\hat{n}_i$ on site $i$~\cite{FN-SF};
(ii) FL (Fermi-like), $\Delta F=\pm 1$, e.g.
the operator $c\dagg_i$ on a site. The two-point Greens function,
the dominant long-range correlation  in a Fermi sea,
belongs with this operator sector.
(iii) SC (superconducting), $\Delta F=\pm 2$; such
operators are the order parameters for superconductivity.
In addition, each operator can be even or odd under exchange
of the ladder's legs, which we denote by appending ``$+$'' or ``$-$''.

\MEMO{SAC: Can this paragraph be cut to save space?}
Generically, the basis operators $\{ \Xop_\nu,  \Yop_\nu \}$ 
do not take the minimal form one would adopt in defining a 
correlation function (even in the {\it free} fermion case).
\OMIT{(where the two-point Greens function
is clearly the slowest decaying and most fundamental correlation.0}
Instead, complicated terms are admixed~\cite{FN-admixture}.
For example, the dominant operator in the FL sector 
not only has single creation operators $c\dagg_i$, but
terms $c\dagg_i \hat{n}_j$.  

\secc{Numerical results and conclusions}
Fig.~\ref{fig:sval-plots}  presents the numerical 
singular values for the CDM in the three limits;
the decay behaviors of the different correlations 
are summarized in Table~\ref{tab:sval-decays},
where they are compared with our knowledge 
from the intervening particle expansion~\cite{SAC-IPE}.
Due to the limited system sizes for ED, the CDM analysis cannot 
determine the dominant kind of correlation at large distances. 
That is practically impossible for Luttinger liquids in any case: 
for the hardcore boson chain 
(related to our models) the asymptotic (superfluid) correlations 
may dominate only after 50-100 sites \cite{troyer}. 
Table~\ref{tab:sval-decays} shows there is a general correspondence
between the decay rate of known correlations and that of the singular
values; the degree of correlation in Fig.~\ref{fig:sval-plots}
tends to be overestimated due to the very small range of $r$.

The rung-fermion case (b) at filling $1/4$ breaks translational symmetry,
with period-2 long-range order.
Examination of Fig.~\ref{fig:sval-plots} (b)
indeed shows the corresponding contrast with the other two cases:
the singular value for the order-parameter operator (CDW$+$) 
is non-decaying and saturates the bound $\sigma=1/2$, whereas 
other kinds of singular values are orders of magnitude smaller.
\MEMO{(Ref.~\cite{SAC-thesis} finds FL behavior for the rung-fermion
case, as expected, when it is doped away from quarter filling.)}

In the boson-pair case (c), as $t'$ grows large
(the boson-pair limit), a crossover is expected 
to asymptotic superconducting (SC) correlations; but 
Fig.~\ref{fig:sval-plots}(c) shows that CDW correlations 
still dominate at all accessible distances, similar
to hardcore bosons~\cite{troyer}.
A partial success the CDM analysis is that the SC
singular values are visibly larger than in the other cases, 
competitive with FL correlations; absent any other knowledge
of this system, the SC order parameter would be flagged for
further study (e.g. analytic, or by quantum Monte Carlo).

\begin{table}
\caption{Correlation behaviors in limiting-case models
Row labels (a, b, c) correspond 
to the panels in Fig.~\ref{fig:sval-plots}.
Columns ``Sim'' summarize
behaviors inferrable from Fig.~\ref{fig:sval-plots}:
``large'', ``medium'', or ``small'' indicate singular values roughly constant 
with $r$, i.e. possible long-range order (values over $10^{-1}$, $10^{-2}$, or 
$10^{-3}$, respectively).
Singular values decaying with $r$ are labeled ``d(fast)'' or ``d(slow)''.
Columns  are labeled by the symmetry sectors as in Fig.~\ref{fig:sval-plots}.
For comparison, the columns ``Th'' are from semi-analytic computations
of Ref.~\onlinecite{SAC-IPE}; exp = exponential decay, 
LRO = long range order.
For the pairing limit (c), the FL correlation exponent 
varies with filling $n$, with $\alpha(n=1/4)\approx 1.1$.
}
\begin{tabular}{|l|cc|cc|cc|cc|}
\hline
  & \multicolumn {8}{c|} {CDM singular values} \\
\cline{2-9}
  & \multicolumn{2}{c|}{ CDW$^+$}   & \multicolumn{2}{c|}{ CDW$^-$} &
    \multicolumn{2}{c|}{ FL$^\pm$}  & \multicolumn{2}{c|}{ SC$^\pm$} \\
\cline{2-9}
  & Sim & Th &  Sim &  Th & Sim & Th &  Sim &  Th \\
\hline  
a & med. & $r^{-2}$ & large & $r^{-1/2}$  & d(slow) & exp & small & $r^{-2.5}$\\
b & large & LRO? & $\sim 0$? & --& d(fast) &  $r^{-1}$ & 0     & $r^{-2.2}$??  \\
c & d(slow)& $r^{-2}$ & med.& $r^{-\alpha}$ & d(slow) & exp & small &$r^{-1/2}$ \\
\hline
\end{tabular}
\label{tab:sval-decays}
\end{table}

In all three cases, most correlations decay generically~\cite{SAC-IPE} as 
$C(r) \sim \cos (2m k_F r+\delta) / |r|^x$,
where $2m k_F$ is an even multiple of the Fermi wavevector and
$x$ is some correlation exponent.
Over a small range of $r$, the with oscillations with $r$ 
obscure the asymptotic $r$ dependence of the singular values.
We conjecture each such correlation is associated with a {\it pair} 
of singular values, oscillating $90^\circ$ out of phase inside the
same envelope.  Ideally, then, one should plot
$\left[\sum'_\nu \sigma_\nu^2\right]^{1/2}$, where ``$\sum'$'' runs 
over just one symmetry sector, to obtain a monotonic decay as $1/|r|^x$.
In practice, for reasons we do not understand, this gave little or no 
improvement.
\OMIT{over the absolute value of the largest singular value from each sector,
which  (for simplicity of analysis) is all we implemented}


To conclude, we have introduced a new tool for analyzing 
exact-diagonalization ground states, using the density
matrix of a pair of clusters to extract {\it all} their
correlations in an unbiased fashion. Furthermore, via a singular-value
decomposition, the kind of operator dominating the correlations 
could be identified, using \eqref{eqn:operatorsvd}.
There are two regimes where asymptotic decays are
not at issue and the correlation density matrix based on exact
diagonalization should be effective.  
First, for systems believed to have
negligible correlations beyond the nearest neighbor
--  e.g.  quantum spin liquids in 
highly frustrated antiferromagnets~\cite{kagome}
-- the CDM is the foolproof way to confirm the absence 
of {\it any} correlations.   
Secondly, in systems having long-range order [such as our case (b)],
the CDM detects the symmetry breaking. 
\OMIT{The two regimes have in common that all correlations 
reach their asymptotic values quickly.}
On the other hand, critical states [such as the Luttinger
liquids of our cases (a) and (c), above] are the {\it least} promising
systems for study by CDM, so long as the system sizes are limited 
by dependence on ED.
But if the CDM and density-matrix renormalization group methods are 
married~\cite{muender-thesis}, the asymptotic scaling may become 
accessible for one-dimensional systems.

Another unbiased method has been proposed to discover the symmetry breaking 
operator from ED using the density matrix~\cite{Fu06}.
It differs from the CDM in two ways:
(i) it is based on the DM of just one cluster;
(ii) it requires not only the ground state's wavefunction,
but that of several low-lying eigenstates which are 
conjectured to be linear combinations of symmetry broken states
(and degenerate in the thermodynamic limit).  
That method is meant only for cases of long-range order,
whereas in principle the CDM identifies the strongest correlations
even in disordered phases.

\OMIT{
(SAC: this issue is too complex for one paragraph).
The problem of detecting the oscillatory dependence,  
and in general the relation of the CDM operators 
at successively larger separations $r$, 
requires an additional analysis involving the
overlap of operators from $\rhoC(r)$ at different separations $r$.
Solving it amounts to finding the scaling operator 
of the renormalization group for our critical state.}

\secc{Acknowledgments}
We thank A. L\"auchli, C. Lhuillier, G. Misguich, 
N. D. Mermin, J. von Delft, and A. Weichselbaum for discussions.
This work was supported by NSF Grant No. DMR-0552461.


\begin{thebibliography}{99}
\bibitem{FN-entangle}
We do not know any property of the CDM that distinguishes
the case of purely classical correlations from that of
quantum entanglement between A and B.
\bibitem{FN-traceXY}
Observe that since $\Tr(\rhoop^{AB})= \Tr(\rhoop^A)=\Tr(\rhoop^B)=1$ is true
of any density matrix, we always have $\Tr(\rhoC)\equiv 0$.
Then since $\Tr_A( \rhoC)  \equiv \Tr_B (\rhoC)=0$, it follows 
that $\Tr \Xop_\nu = \Tr \Yop_\nu =0$ 
for each of the SVD operators.
\bibitem{FN-rhoCbound}
Eq.~\eqref{eqn:rhoCbound} follows because 
(using \eqref{eqn:frob-norm} and \eqref{eqn:frob-innerprod})
$(\rhoop^{AB},\rhoop^{A}\otimes\rhoop^{B})_F=
\langle \rhoop^A \otimes \rhoop^B \rangle =
\langle \rhoop^A \rangle \langle \rhoop^B \rangle =
\|\rhoop^A \|^2 \|\rhoop^B \|^2.$
We achieve the bound when $\rhoop^{AB}$ describes a
pure state (norm 1), while every 
eigenvalue of $\rhoop^{A}$ (and necessarily of $\rhoop^{B}$) 
is $1/D$, i.e., when $A\cup B$ is in a maximally entangled state.
\bibitem{FN-restoresymm}
Our density matrices are averaged over 
degenerate ground states, if any,
to restore all symmetries,
as explained in Ref.~\onlinecite{SAC-interacting}.
\end{thebibliography}
\end{document}